\pdfoutput=1
\documentclass[aps,prd,amsmath,floats,floatfix, twocolumn,
superscriptaddress,nofootinbib,showpacs,longbibliography]{revtex4-1}

\usepackage[T1]{fontenc}
\usepackage[utf8]{inputenc}
\usepackage{lmodern}
\usepackage{color,soul}
\usepackage{verbatim}
\usepackage{dsfont}

\usepackage[dvipsnames, usenames]{xcolor}
\definecolor{linkcolor}{rgb}{0.0,0.3,0.5}
\usepackage[hypertexnames=false, unicode, colorlinks=true, linkcolor=linkcolor,
citecolor=linkcolor, filecolor=linkcolor,urlcolor=linkcolor,
pdfusetitle]{hyperref}

\usepackage[all]{hypcap}
\usepackage{graphicx}
\usepackage{xspace}
\usepackage{amssymb}
\usepackage[normalem]{ulem} 
\usepackage{bm} 

\usepackage{microtype}

\usepackage[english]{babel}
\usepackage{blindtext}

\usepackage{xcolor}   
\newcommand{\rev}{}
\newcommand{\revv}{}
\newcommand{\revvv}{}

\graphicspath{%
  {figs/}%
}

\DeclareMathAlphabet{\mathpzc}{OT1}{pzc}{m}{it}

\newcommand{\mrm}{\mathrm }

\newcommand{\D}{\mathrm{d}}

\DeclareMathOperator{\diag}{diag}
\DeclareMathOperator{\sign}{sign}

\begin{document}

\title{Time-energy uncertainty relation from subcycle mode \\ vacuum fluctuations of a quantum field}

\newcommand{\UQ}{\affiliation{Centre for Quantum Computation and Communication Technology, School of Mathematics and Physics, The University of Queensland, St. Lucia, Queensland, 4072, Australia}}

\author{Achintya Sajeendran}
\email{a.sajeendran@uq.edu.au}

\author{Timothy C. Ralph}
\email{ralph@physics.uq.edu.au}

\UQ 
\hypersetup{pdfauthor={Sajeendran et al.}}

\date{\today}

\begin{abstract}
   The time-energy uncertainty relation is often invoked as a heuristic explanation for virtual particles in interacting quantum field theories. However, this interpretation breaks down upon closer scrutiny for several reasons, particularly since virtual particles do not have a well-defined temporal extension. Although concrete derivations and interpretations of time-energy uncertainty bounds in quantum mechanics have been established, most famously by Mandelstam and Tamm in 1945, there is no known rigorous connection between these bounds and the concept of virtual particles in quantum field theory. In this work, we use a model in which the vacuum particle content associated with subcycle, spatiotemporally localised modes of a free scalar field can be converted into excitations of a rapidly-switched harmonic-oscillator Unruh-DeWitt detector coupled to the conjugate field. Defining the time uncertainty as the effective duration of the detector-field interaction and identifying the contribution to the energy fluctuations of the detector resulting from the subcycle mode vacuum fluctuations, we show that a time-energy uncertainty relation is satisfied in the deep subcycle regime. Our results provide a concrete operational meaning to the textbook heuristic picture of virtual particles in quantum field theory in terms of the time-energy uncertainty principle.
\end{abstract}

\maketitle

\textit{Introduction}---One of the essential properties of quantum mechanics is the existence of a fundamental theoretical limit on the precision with which one can resolve a pair of incompatible (non-commuting) observables. Mathematically, given two incompatible observables corresponding to self-adjoint operators $\hat A$ and $\hat B$, the uncertainties in these observables, $\Delta A$ and $\Delta B$, must obey the Heisenberg--Robertson (HR) uncertainty bound \cite{heisenberg1927anschaulichen,robertson1929uncertainty}.
In particular, the famous Heisenberg uncertainty principle (UP) is obtained by considering such a bound for the position and momentum observables of a particle, resulting in $\Delta x\Delta p\geq \frac{\hbar}{2}$. Classical physics possesses no such limit---all uncertainties in observables measured in experiment can be attributed to measurement errors and external factors, and knowledge of the value of one observable does not preclude knowledge of another.

Based on a heuristic argument, Heisenberg \cite{heisenberg1927anschaulichen} further suggested an uncertainty relation between time and energy. His original derivation of the Heisenberg UP was based on noting that the position-space wavefunction $\psi(x)$ and the momentum-space wavefunction $\phi(p)$ are related by Fourier transform. Similarly, time and energy are Fourier conjugate variables, leading to the heuristic time-energy uncertainty relation,
\begin{align}
    \Delta E\Delta t \geq \frac{\hbar}{2}.\label{eq1 standard TEUR}
\end{align}
However, since time is not treated as a self-adjoint observable in quantum mechanics, this bound cannot be considered a special case of the HR uncertainty relation. The interpretation of time as an external parameter is forced on standard quantum theory, as the existence of a self-adjoint time operator $\hat T$ that is canonically conjugate to the Hamiltonian $\hat H$, $[\hat T,\hat H]=i\hbar$, would imply that $\hat H$ is unbounded. Nevertheless, as first shown by Mandelstam and Tamm \cite{mandelstam1945uncertainty} in 1945, the time-energy uncertainty relation of Eq. \eqref{eq1 standard TEUR} can be derived rigorously given an appropriate definition of the `time uncertainty' $\Delta t$, with energy uncertainty defined simply as the uncertainty in the Hamiltonian, $\Delta H$, of the system. Physically, in the Mandelstam--Tamm (MT) prescription, $\Delta t$ is defined to be the time taken for the expectation value of a self-adjoint observable, $\langle \hat A\rangle$ to change by one standard deviation, $\Delta A$. Mathematically, 
\begin{align}
    \Delta t &\equiv \frac{\Delta A}{\left| \frac{\D\langle \hat A \rangle}{\D t} \right|}.\label{eq6 mandelstam-tamm time def}
\end{align}

Furthermore, there exist bounds on the time taken for a system to evolve into a state orthogonal to its initial state $|\psi\rangle$, known as quantum speed limits (QSLs) \cite{mandelstam1945uncertainty,margolus1998maximum}. The MT uncertainty relation \cite{mandelstam1945uncertainty}, as well as these QSL bounds \cite{mandelstam1945uncertainty,margolus1998maximum}, demonstrate that it is possible to rigorously derive bounds involving the resolution of time and energy in standard quantum theory, regardless of the non-existence of a self-adjoint time operator.

If the particle interpretation of virtual excitations is taken seriously, virtual particles correspond to (generally) off-shell excitations of the field. Heuristically, this is explained using the time-energy uncertainty relation (TEUR): virtual particles are excitations of the field that violate energy conservation while existing for only a very short duration. However, it is clear that this interpretation breaks down under further scrutiny. Namely, in any QFT possessing time-translation symmetry, energy is conserved at each vertex of the Feynman diagram. Internal lines really correspond to Feynman propagators, which are off-shell objects, rather than excitations of the field. Most problematically, virtual particles do not even have a well-defined temporal extension. The heuristic interpretation of virtual particles in terms of TEUR is therefore not to be taken seriously. However, this interpretation can have meaning if we replace virtual particles with temporally localised modes.

In this Letter, we show how this interpretation can be modified, leading to a concrete operational interpretation of the particle content associated with Gaussian subcycle modes of a quantum field in terms of TEUR. Namely, using a model employed by Onoe \emph{et al.} \cite{onoe2022realizing}, the particle content in the field vacuum associated with these Gaussian subcycle modes can be converted into excitations of a probe system coupled to the field over a short timescale. \rev{Although it is well-known that particle detectors coupled to quantum fields over a finite time will exhibit excitations \cite{sriramkumar1996finite,higuchi1993uniformly,sriramkumar1994response,louko2006often,satz2007then,shevchenko2017finite,stargen2026finite}, none of these approaches establish a concrete relationship between the nontrivial finite-time response and the virtual particle interpretation of the time-energy uncertainty principle, although this is sometimes taken as a loose heuristic interpretation without further justification (see for example Ref. \cite{sriramkumar1996finite}). \revvv{In our approach, we are able to calculate an approximation of the energy uncertainty of the probe system, identifying the contribution, $\Delta E$, to this uncertainty associated with the subcycle mode vacuum fluctuations.} Defining $\Delta t$ to be the effective duration of the interaction, and assuming the idealised regime of unit-efficiency we derive a time-energy uncertainty relation in the deep subcycle regime, showing that \revv{$\Delta E\Delta t= \frac{\hbar}{\sqrt{2\pi}}$. Although the relation we derive does not provide a fundamental lower bound on the time-energy uncertainty product of all quantum states unlike standard uncertainty relations, it still possesses the form $\Delta E\Delta t\sim \hbar$, thereby capturing the essential features of the time-energy uncertainty principle, insofar as short-lived excitations have energy uncertainties reciprocal to their lifetimes. Indeed, this is more compatible with the heuristic argument for virtual particles in terms of the time-energy uncertainty principle. Our work provides the first concrete operational basis for this interpretation.}}

Unless otherwise stated, we use units in which $\hbar=c=1$, and we use the convention $\eta_{\mu\nu}=\diag(-1,+1,+1,+1)$ for the Minkowski metric.

\textit{Gaussian subcycle modes of a quantum field}---We consider for simplicity a real massless scalar field in $(3+1)$-dimensional Minkowski spacetime. We will further assume that the modes of interest propagate in one direction, $x$, which is, for example, a physically relevant regime for fields propagating in an optical fiber \cite{blow1990continuum}. While we consider a field in free space in this work, we will eventually couple the field to a particle detector with finite spatial extension in the directions transverse to these modes, in which case we can make the same approximation.

The right-moving sector of such a field can be written as
\begin{align}
    \hat\phi(t,x) &= \int_{-\infty}^\infty \D\omega\:u_\omega(t,x)\hat a_\omega, \\
    u_\omega(t,x) &= \sqrt{\frac{1}{4\pi|\omega|A}}e^{-i\omega(t-x)}.
\end{align}
Here, $A$ is simply an effective factor with units of area which, in our case, can be interpreted as the transverse area of a particle detector, as we show below. Note also that we use the convention $\hat a_{-\omega}=\hat a_\omega^\dagger$, so that the mode operators satisfy the commutation relation $[\hat a_\omega,\hat a_{\omega'}^\dagger]=\sign(\omega)\delta(\omega-\omega')$.

Similarly, we can write
\begin{align}
    \hat\pi(t,x) &= \int_{-\infty}^\infty \D\omega\:v_\omega(t,x)\hat a_\omega,\\
    v_\omega(t,x) &= -\sign(\omega)i\sqrt{\frac{|\omega|}{4\pi A}}e^{-i\omega(t-x)}.
\end{align}

Now, suppose we decompose the field into a complete orthonormal set of wavepacket modes \cite{rohde2007spectral}. Suppose that one of these modes is a Gaussian pulse of the form
\begin{align}
    u_g(t,x) &= \frac{1}{(2\pi)^{1/4}}\sqrt{\frac{\sigma}{A\omega_0}}e^{-\sigma^2(t-x-t_0)^2-i(t-x)\omega_0}.\label{eq7 Gaussian mode function}
\end{align}
Here, the Gaussian pulse $u_g(t,x)$ has carrier frequency $\omega_0/(2\pi)$ and temporal variance $1/(2\sigma^2)$. The position of its amplitude maximum crosses $x=0$ at time $t=t_0$. We consider such a Gaussian mode to be in the subcycle regime when $2\pi\omega_0^{-1}\gg \sigma^{-1}$, i.e. one period of carrier frequency is much longer than a single standard deviation of the Gaussian envelope.

The corresponding mode operator which annihilates the Gaussian mode $u_g(t,x)$ is given by 
\begin{align}
    \hat a_g &= \int_{-\infty}^\infty \D\omega\:f_g(\omega)\hat a_\omega, \\
    f_g(\omega) &= \frac{1}{(2\pi)^{1/4}}\sign(\omega)\sqrt{\frac{|\omega|}{\omega_0\sigma}}e^{-it_0(\omega-\omega_0)-\frac{(\omega-\omega_0)^2}{4\sigma^2}}.\label{eq9 Gaussian spectral coeffs}
\end{align}
It can be shown that the average particle number of these Gaussian modes in the field vacuum is given by (see Ref. \cite{onoe2022realizing} for the definition of $\theta_g$)
\begin{align}
    \langle \hat n_g \rangle &= \int_0^\infty \D\omega\:|f_g(-\omega)|^2=\sinh^2(\theta_g),\label{eq10:average particle number}
\end{align}
where $\hat n_g\equiv \hat a_g^\dagger\hat a_g$ and $\langle \cdot \rangle$ denotes expectation value in the field vacuum. We note that the expected particle content \eqref{eq10:average particle number} is generally nonzero for any Gaussian mode. In the narrowband regime ($\omega_0\gg \sigma$), the overlap of the mode function \eqref{eq9 Gaussian spectral coeffs} into the negative frequency domain will be sufficiently small so that $\langle \hat n_g \rangle \approx 0$. However, in the subcycle regime, the overlap into negative frequencies can become appreciably large, meaning that the nonzero expected particle content in the vacuum cannot be neglected. We note that there is no contradiction here, since spatiotemporally localised modes can be excited, even if the field is in the vacuum state defined with respect to the normal modes. In the next section, we show how this subcycle vacuum particle content can be converted into excitations of a mode-selective probe system.

\textit{Mode-selective Unruh-DeWitt detection model}---In general, the nonzero expectation value of a number operator cannot immediately be assumed to possess any concrete physical meaning. It is important to assign an operational meaning to this quantity, for example, by probing these excitations using a particle detector in some physical regime \cite{birrell1984quantum,unruh1984happens,takagi1986vacuum,sriramkumar2002probes,wald1994quantum}. For mathematical simplicity, we will use a simplified model given by a harmonic oscillator Unruh-DeWitt (UDW) detector that is coupled linearly to the conjugate field $\hat\pi$. This has been shown to serve as a useful toy model for a realistic scheme using electro-optic sampling \cite{onoe2022realizing}.

The harmonic oscillator UDW detector couples to the (conjugate) field via the interaction-picture interaction Hamiltonian,
\begin{align}
    \hat H_I(\tau) &= A\lambda\chi(\tau)\hat Q(\tau)\hat\pi\left( t(\tau),x(\tau) \right),\label{eq11 interaction Hamiltonian} \\
    \hat Q(\tau) &= \sqrt{\frac{1}{2A}}\left( \hat u e^{-i\omega_u \tau}+\hat u^\dagger e^{i\omega_u\tau} \right),\label{eq38:effective monopole moment}
\end{align}
where $\hat\pi\left( t(\tau),x(\tau) \right)$ is the conjugate momentum field pulled back to the worldline of the detector. The area factor $A$ now takes on an operational interpretation as the cross-sectional area of the detector in the transverse directions (see Appendix \ref{sec:appendix a}). Here, $\lambda$ is the maximum coupling strength between the detector and field, and $\chi(\tau)$ is the switching function, which controls the effective duration of the interaction. The quadrature operator $\hat Q(\tau)$ is the monopole moment of the detector. For simplicity, we will consider a detector on the inertial worldline $\left( t(\tau),x(\tau) \right)=(\tau,0)$.

We specifically consider the case of narrow Gaussian switching function, $\chi(t)=e^{-\sigma_u^2(t-t_u)^2}$, corresponding to a rapid detector-field interaction ($\sigma_u\gg \omega_u$). \revvv{We work in an operational regime under which the effects of time-ordering can be neglected \cite{christ2013theory,quesada2014effects,lipfert2018bloch}, giving the effective interaction-picture evolution unitary},
\begin{align}
    \hat U_I &\approx \exp\left( -i\int_{-\infty}^{\infty} \D\tau\:\hat H_I(\tau) \right).\label{eq:effective subcycle unitary}
\end{align}

It can be shown that (see Ref. \cite{onoe2022realizing}) if we set $\omega_u=\omega_0$, $\sigma_u=\sigma$, and $t_u=t_0$, this reduces to a beamsplitter-type unitary,
\begin{align}
    \hat U_I &\approx \exp\left[ \theta_u(\hat a_g\hat u^\dagger-\mrm{H.c.}) \right],\label{eq14:interaction picture unitary first-order approx}
\end{align}
where $\theta_u \equiv -\frac{\lambda}{2}\sqrt{\frac{\omega_0}{\sigma}}\left( \frac{\pi}{2} \right)^{1/4}$. In particular, the detector effectively only couples to the Gaussian subcycle mode $\hat a_g$, acting as a mode-selective probe.

\textit{Deriving the time-energy uncertainty relation}---After the interaction, the Heisenberg picture detector annihilation operator evolves to
\begin{align}
    \hat u' &= \hat U_I^\dagger\hat u\hat U_I=\cos(\theta_u)\hat u+\sin(\theta_u)\hat a_g.
\end{align}
\revvv{Assuming that the detector's initial state (at $t=-\infty$) is its ground state, after the interaction (at $t=\infty$), its average excitation number is given by
\begin{align}
    \langle \hat n' \rangle &= \langle 0|\langle 0_\mrm{D}|\hat u'^\dagger\hat u'|0_\mrm{D}\rangle|0\rangle=\sin^2(\theta_u)\sinh^2(\theta_g),\label{eq16 average detector excitation number}
\end{align}
where $|0_\mrm{D}\rangle$ is the ground state of the detector and $|0\rangle$ is the field vacuum. Since $\langle \hat n_g\rangle=\sinh^2(\theta_g)$, we identify $\eta_u=\sin^2(\theta_u)$, where $\eta_u$ is the effective efficiency of the conversion of the mean subcycle mode vacuum particle content into detector excitations. We will consider the idealised regime in which $\theta_u=\frac{\pi}{2}$, or $\eta_u=1$, corresponding to a unit-efficiency conversion. To achieve this regime, we require the coupling constant to be $\lambda=-(2\pi^3)^{1/4}\sqrt{\frac{\sigma}{\omega_0}}$, which becomes large in the deep subcycle regime $\sigma \gg \omega_0$. While this strong-coupling regime would be very difficult to achieve in practice, we consider this regime as an operational idealisation in which all of the vacuum subcycle particle content is converted into detector excitations. We treat the finite-efficiency (weak-coupling) case, which could potentially be realised in electro-optic sampling experiments \cite{onoe2022realizing}, in Appendix \ref{sec:appendix b}.

For $\theta_u=\frac{\pi}{2}$, it is readily shown using Wick's theorem that (see Appendix \ref{sec:appendix c})
\begin{align}
    (\Delta n')^2 &= (\Delta n_g)^2,\label{eq17 number variance for unit efficiency} \\
    &= \sinh^2(\theta_g)\cosh^2(\theta_g)+\frac{\sigma^2}{2\pi\omega_0^2}e^{-\omega_0^2/\sigma^2}.
\end{align}
Thus, the post-interaction energy variance of the detector is 
\begin{align}
    (\Delta E)^2 &= \omega_0^2\left( \sinh^2(\theta_g)\cosh^2(\theta_g)+\frac{\sigma^2}{2\pi\omega_0^2}e^{-\omega_0^2/\sigma^2} \right).
\end{align}
Defining $\Delta t=\frac{1}{\sqrt{2}\sigma}$ to be the standard deviation of the normalised switching function $\chi(t)$ (which also coincides with the temperal extension of the Gaussian mode annihilated by $\hat a_g$), we get 
\begin{align}
    (\Delta E)^2 (\Delta t)^2 &= \frac{\omega_0^2}{2\sigma^2}\sinh^2(\theta_g)\cosh^2(\theta_g)+\frac{e^{-\omega_0^2/\sigma^2}}{4\pi}.
\end{align}
We note that the appropriate limiting regime to consider is $\frac{\omega_0}{\sigma}\ll 1$. Firstly, the subcycle regime already requires that $\omega_0\ll \sigma$ by definition. We note that the subcycle regime is crucial for our treatment, as it coincides with the rapid-switching regime in which the effective duration of the detector-field interaction is short. Under this condition, and in the operational regime of the electro-optic sampling implementation of this detection model considered in Ref. \cite{onoe2022realizing}, the effects of time-ordering on the dynamics of the statistical moments of interest can safely be neglected \cite{quesada2014effects,christ2013theory,lipfert2018bloch,onoe2022realizing} as we have done in Eq. \eqref{eq:effective subcycle unitary}. Furthermore, if $\omega_0$ is comparable to $\sigma$, then the wavepacket will contain excitations that are far from integer multiples of the detector resonance, $\omega_0$, and will therefore have a lower probability of exciting the detector. In the regime $\omega_0\ll\sigma$, the energy gap becomes narrow, allowing a complete exchange of vacuum subcycle particle content into detector excitations.} In this regime, we obtain the following relation between the time and energy uncertainties:
\begin{align}
    \Delta E\Delta t &= \frac{\hbar}{\sqrt{2\pi}},\label{eq22 time-energy uncertainty relation}
\end{align}
where we have restored factors of $\hbar$.

\textit{Discussion}---We note that the right-hand side of the relation \eqref{eq22 time-energy uncertainty relation} is slightly less than the usual lower bound, $\hbar/2$, required by the Mandelstam--Tamm relation \cite{mandelstam1945uncertainty}. However, in the derivation of our relation \eqref{eq22 time-energy uncertainty relation}, we have not defined $\Delta t$ with respect to the evolution of a self-adjoint observable, as in Eq. \eqref{eq6 mandelstam-tamm time def}; rather, we have defined it as an effective duration of the detector-field interaction, which coincides with the temporal extension of the Gaussian envelope \eqref{eq7 Gaussian mode function}. This definition of $\Delta t$ is a formalisation of the typical heuristic interpretation of $\Delta t$ as the `lifetime' of a virtual excitation, and is therefore more compatible with our purposes than the Mandelstam--Tamm prescription. As such, this apparent inconsistency should not be considered a genuine violation.

Furthermore, notice that Eq. \eqref{eq22 time-energy uncertainty relation} is not in the form of an inequality. Our relation, unlike that derived by Mandelstam and Tamm \cite{mandelstam1945uncertainty}, is not to be interpreted as a generic lower bound on the time-energy uncertainty product of arbitrary quantum states. Rather, the relation \eqref{eq22 time-energy uncertainty relation} still exhibits the essential features of the time-energy uncertainty principle as a relationship between conjugate variables, $\Delta E\Delta t\sim\hbar$, in the sense that short-lived excitations possess an energy uncertainty inversely proportional to their lifetime. The Mandelstam--Tamm relation, in contrast, does not possess such an interpretation in terms of the energy uncertainty of finite-time excitations of a quantum field.

In this work, we have given a concrete operational meaning for the typical heuristic argument for virtual particles in QFT. For several reasons, this heuristic argument has long been known to be problematic. Crucially, virtual particles do not possess a well-defined temporal extension. We therefore argue that it is more appropriate to replace virtual particles in this heuristic argument with spatiotemporally localised modes. In particular, we have considered Gaussian subcycle modes, identifying an idealised particle detector model in which the subcycle particle content in the field vacuum is converted into detector excitations with unit efficiency. \revvv{We note that the qualitative aspects of our conclusions are unaffected in the more realistic and experimentally relevant finite-efficiency case, which we treat in Appendix \ref{sec:appendix b}.} We showed that the energy uncertainty, $\Delta E$, of the detector resulting from subcycle mode fluctuations in the field vacuum is inversely proportional to the effective interaction time $\Delta t$ in the deep subcycle regime. \revvv{Importantly, our result could potentially be tested in realisable electro-optic sampling experiments \cite{onoe2022realizing}, accounting for the finite-efficiency (weak coupling) conversion in such setups (see Appendix \ref{sec:appendix b}).}

We note that, although we have specialised to subcycle Gaussian-switched UDW detectors in this Letter, our approach may be generalisable to arbitrary temporally-localised interactions that induce squeezing (see for example the setup of Ref. \cite{rukan2025truncated}). This Letter lays the groundwork for a general operational approach to time-energy uncertainty relations inspired by the heuristic argument for virtual particles, thereby providing a complementary perspective to the Mandelstam--Tamm approach. Furthermore, since we consider a detector with finite cross-sectional area, we expect that there will be nontrivial corrections to our relation in general curved spacetimes. This may provide further insight into gravitational effects on quantum uncertainty, giving a complementary, operationally grounded direction to Generalised Uncertainty Principles \cite{witten1996reflections,gross1988string,kempf2009information,bojowald2012generalized,scardigli1999generalized,amelino2011principle,amelino2002doubly}---the latter of which are essentially based on modified commutation relations. Our present work will serve as the basis for such future developments in these directions, \revvv{aiming to bridge foundational aspects of quantum uncertainty, the interplay between quantum theory and general relativity, and potentially realisable quantum optics experiments.}

\begin{acknowledgments}
    We thank Magdalena Zych, Fabio Costa, Joshua Foo, and Sara Butler for useful discussions. This research was supported by the Australian Research Council Centre of Excellence for Quantum Computation and Communication Technology (Project No. CE170100012).
\end{acknowledgments}

\appendix 

\begin{widetext}
\section*{End Matter}

\section{Cross-sectional smearing of an Unruh-DeWitt detector}\label{sec:appendix a}

In this section, we show that the effective area factor $A$ can be interpreted as the transverse extent of the UDW detector. Consider for simplicity a static UDW detector, interacting with the conjugate field via the interaction-picture interaction Hamiltonian,
\begin{align}
    \hat H_I(t) &= \lambda\chi(t)\hat Q(t)\int \D^2\mathbf{x}_\perp\:g(\mathbf{x}_\perp)\hat\pi(t,0).
\end{align}
Here, $g(\mathbf{x}_\perp)\in [0,1]$ is the cross-sectional smearing function of the detector. The detector is taken to be pointlike in the $x$ direction, which is the direction in which the field modes of interest propagate.

Assuming that $g$ has support only within a cylindrical region $R$, we may approximate
\begin{align}
    \hat H_I(t) &\approx A\lambda\chi(t)\hat Q(t)\hat\pi(t,0).
\end{align}
Furthermore, for the action associated with this interaction to be dimensionless, we require that $\hat H_I$ has mass dimension one (in units $\hbar=c=1$), i.e. $[\hat H_I]=1$. We thus have
\begin{align}
    1 &= [A]+[\lambda]+[\hat Q(t)]+[\hat\pi(t,0)] = [\lambda]+[\hat Q(t)].
\end{align}
Since we choose $[\lambda]=0$, we require $[\hat Q(t)]=1$. This is why we introduce the additional factor of $1/\sqrt{A}$ in the monopole moment \eqref{eq38:effective monopole moment}.

\revvv{\section{Finite-efficiency coupling regime}\label{sec:appendix b}

In this section, we treat the finite-efficiency regime, corresponding to the weak-coupling regime relevant for electro-optic sampling implementations of a rapidly-switched UDW detector \cite{onoe2022realizing}. We note that, even in the finite-efficiency case, the contribution to the number fluctuations in the detector resulting from the subcycle mode field vacuum fluctuations can be clearly identified.

For $\theta_u\neq \frac{\pi}{2}$ ($\eta_u\neq 1$), the post-interaction number variance of the detector is given by
\begin{align}
    (\Delta n')^2 &= \sin^4(\theta_u)(\Delta n_g)^2+\sin^2(\theta_u)\cos^2(\theta_u)\langle \hat n_g \rangle, \\
    &= \eta_u^2 (\Delta n_g)^2+\eta_u(1-\eta_u)\langle \hat n_g \rangle.
\end{align}
Note that this reduces to Eq. \eqref{eq17 number variance for unit efficiency} for $\eta_u=1$. We identify the first term as the contribution resulting from the subcycle vacuum fluctuations, while the second term is due to the noise produced by mixing the subcycle mode and detector ground state fluctuations with imperfect efficiency ($\eta_u<1$, $\theta_u<\frac{\pi}{2}$). The corresponding post-interaction energy variance of the detector can be written as
\begin{align}
    (\Delta E')^2 &= \omega_0^2\left[ \eta_u^2(\Delta n_g)^2+\eta_u(1-\eta_u)\langle \hat n_g\rangle \right].
\end{align}
We are interested specifically in the contribution to this energy variance due to the subcycle vacuum fluctuations:
\begin{align}
    (\Delta E_\mrm{sub})^2 &= \omega_0^2\frac{(\Delta E')^2-\eta_u(1-\eta_u)\langle \hat n_g\rangle}{\eta_u^2},\label{eq subcycle part of energy fluctuations} \\ &\equiv \omega_0^2(\Delta n_\mrm{sub})^2.
\end{align}
Indeed, $\lambda$, $\omega_0$, and $\sigma$ would be controllable parameters in any realisable electro-optic sampling experiment \cite{onoe2022realizing}, and they completely determine the values of $\eta_u$ and $\langle \hat n_g\rangle$ (up to external noise factors and loss). Therefore, after obtaining $(\Delta E')^2$ from multiple measurements, the contribution to this due to fluctuations in the vacuum subcycle-mode particle content can be inferred after adjusting according to Eq. \eqref{eq subcycle part of energy fluctuations}.

We find that
\begin{align}
    (\Delta E_\mrm{sub})^2 &= \omega_0^2\left( \sinh^2(\theta_g)\cosh^2(\theta_g)+\frac{\sigma^2}{2\pi\omega_0^2}e^{-\omega_0^2/\sigma^2} \right).
\end{align}
After defining $\Delta t=\frac{1}{\sqrt{2}\sigma}$ as we did in the idealised unit-efficiency case, we obtain the following time-energy uncertainty relation in the regime $\omega_0\ll\sigma$:
\begin{align}
    \Delta E_\mrm{sub}\Delta t \approx \frac{\hbar}{\sqrt{2\pi}},
\end{align}
where we have restored factors of $\hbar$.

}

\section{Computing the second moment of the Gaussian mode number operator}\label{sec:appendix c}

In this section, we will calculate the second moment of the Gaussian mode number operator, $\langle \hat n_g^2 \rangle$. By Wick's theorem, we have
\begin{align}
    \langle \hat n_g^2\rangle &= \langle \hat a_g^\dagger\hat a_g\hat a_g^\dagger\hat a_g \rangle, \\
    &= |m_g|^2+2n_g^2+n_g,\label{eqC2}
\end{align}
where $m_g\equiv \langle \hat a_g^2\rangle$ and $n_g\equiv \langle \hat a_g^\dagger\hat a_g\rangle=\sinh^2(\theta_g)$ (see Eq. \eqref{eq10:average particle number}). We compute
\begin{align}
    m_g &= \langle \hat a_g^2\rangle, \\
    &= \int_{-\infty}^\infty \D\omega\int_{-\infty}^\infty \D\omega'\:f_g(\omega)f_g(\omega')\langle \hat a_\omega\hat a_{\omega'}\rangle, \\
    &= \int_{0}^\infty \D\omega\int_{-\infty}^0 \D\omega'\:f_g(\omega)f_g(\omega')\langle \hat a_\omega\hat a_{\omega'}\rangle, \\
    &= \int_0^\infty \D\omega\int_0^\infty \D\omega'\:f_g(\omega)f_g(-\omega')\langle \hat a_\omega\hat a_{\omega'}^\dagger \rangle, \\
    &= \int_0^\infty \D\omega\:f_g(\omega)f_g(-\omega), \\
    &= -\frac{\sigma}{\sqrt{2\pi}\omega_0}e^{-\omega_0^2/(2\sigma^2)}.
\end{align}
Using Eq. \eqref{eqC2}, we obtain
\begin{align}
    \langle \hat n_g^2\rangle &= n_g(1+2n_g)+|m_g|^2, \\
    &= \sinh(\theta_g)\left( 1+2\sinh^2(\theta_g) \right)+\frac{\sigma^2}{2\pi\omega_0^2}e^{-\omega_0^2/\sigma^2}, \\
    &= \sinh(\theta_g)\left( \sinh^2(\theta_g)+\cosh^2(\theta_g) \right)+\frac{\sigma^2}{2\pi\omega_0^2}e^{-\omega_0^2/\sigma^2}, \\
    \langle \hat n_g^2\rangle &= \sinh^4(\theta_g)+\sinh^2(\theta_g)\cosh^2(\theta_g)+\frac{\sigma^2}{2\pi\omega_0^2}e^{-\omega_0^2/\sigma^2},
\end{align}
as desired.

\end{widetext}

\bibliography{ref}

\end{document}